\documentclass[aps,twocolumn]{revtex4}
\usepackage{graphicx}   
\usepackage{latexsym}   

\newcommand{\bld}[1]{\mbox{\boldmath $#1$}}     
\newcommand{\etal}{{\em et al.}}		
\newcommand{\ie}{{\em i.e.}}			
\newcommand{\fm}{{\rm fm}}			
\newcommand{\MeV}{{\rm MeV}}			

\newcommand{\beq}{\begin{equation}}
\newcommand{\eeq}{\end{equation}}
\newcommand{\beqar}{\begin{eqnarray}}
\newcommand{\eeqar}{\end{eqnarray}}

\newcommand{\half}{\mbox{{$1\over2$}}}

\newcommand{\del}{\partial}

\newcommand{\h}{h}
\newcommand{\SKIP}[1]{{ }}

\begin{document}

\title{Catapult neutrons from neck snapping in fission}
\author{J\o rgen Randrup$^1$, Roberto Capote$^{2,3}$, and Ramona Vogt$^{4,5}$}

\affiliation{
  $^1$Nuclear Science Division, Lawrence Berkeley National Laboratory,  
  Berkeley, CA 94720, USA \break
  $^2$NAPC, Nuclear Data Section, International Atomic Energy Agency, 
  1220 Vienna, Austria \break
  $^3$Suncoast Data Evaluation, Miami, FL 33154, USA\break
  $^4$Physics Division, Lawrence Livermore National Laboratory, 
  Livermore, CA 94551, USA \break
  $^5$Physics Department, University of California, Davis, 
  CA 95616, USA}

\date{April 12, 2026}

\begin{abstract}
Dynamical fission calculations show that the post-scission configurations
resemble two colinear pear-shaped fragments
whose juxtaposed surface bulges subside relatively quickly,
as the fragments acquire smoother shapes.
The associated rapid speed of the healing bulge surface
may boost nucleons in the fragment to energies sufficient for emission.
The present study explores this mechanism by following the fate
of nucleons that are reflected off the inwards moving bulge surface.
The simulations suggest that the mechanism may produce high-energy neutrons
at the level of a few per cent.
\end{abstract}

\maketitle

\section{Introduction}

A variety of experiments over the years have reported some evidence
of fission neutrons with energies far in excess of those typical of evaporation.
Because such neutrons could not arise from the equilibrated fragments, 
they have been associated with the scission process and, therefore,
they are collectively denoted as ``scission neutrons'',
although the mechanism behind their production has not yet been established.


Shortly after the experimental evidence for fission 
was reported \cite{Hahn1939} and interpreted \cite{Meitner1939},
it was expected, and experimentally verified,
that fission fragments are radioactive and emit delayed neutrons 
\cite{McMillanPR55,RobertsPR55}.
Subsequently, Anderson \etal\ \cite{AndersonPR55}
not only reported evidence of additional neutrons
produced on a much shorter time scale than the weak decays
but also suggested that these prompt fisssion neutrons 
may arise either from the highly excited fragments 
or are produced during the division process itself.
At the same time, independently, Szilard \etal\ \cite{SzilardPR55}
also found evidence for prompt neutrons.
A few months later, von Halban \etal\ \cite{Halban1939c}
reported measurements of the prompt neutrons
produced in thermal fission of uranium, employing two different methods.
The first method used a threshold reaction 
(nowadays called a neutron dosimetry reaction), $^{32}$S($n$,$p$)$^{32}$P,
to show the existence of fission neutrons with energy above 2\,\MeV,
while the second method directly measured the neutron energy distribution,
finding neutrons with energies greater than 11 MeV.

During the Manhattan project Feather carried out a first quantitative analysis 
of fission neutrons \cite{Feather1942},
as discussed in a recent historical review \cite{Chadwick2023}.
Of particular relevance from that era are the measurements by Richards \etal\
\cite{Richards1944a,Richards1944b}, 
who used photographic plates to detect neutrons with energies above 9 MeV 
for $^{235}$U($n_{\rm th}$,f) and $^{239}$Pu($n_{\rm th}$,f).
Later on Watt \cite{Watt1952} extended the measured prompt neutron spectrum 
up to 17\,\MeV\ using a proton recoil counter.

An analysis by Bowman \etal\ \cite{BowmanPR126}
of the differential distribution of prompt neutrons in $^{252}$Cf(sf)
found that the evaporation contribution from the accelerated fragments
must be supplemented by a more energetic component of $\approx$10\% 
centered in the frame of the fissioning system.
A similar conclusion was reported soon afterwards by Kapoor \etal\
\cite{KapoorPR131} who found that this harder component had an evaporation-like
spctrum and speculated that it might arise from evaporation during the
evolution from saddle to scisison. 
At the same time, Skarsv{\aa}g and Bergheim \cite{SkarsvaagNPA45}
reported that about 15\% of the neutrons had an evaporation-like spectrum
in the CM system with a higher temperature.
However, ten years later Skarsv{\aa}g \cite{SkarsvaagPS7}
 reported that a reanalysis of the data
showed that those neutrons arose from neutron evaporation 
during the acccleration stage of the fragments and concluded that
no convincing evidence for scission neutrons had been found.

Subsequently, analysis of measured $n$-$n$ angular correlations
by Pringle and Brooks \cite{PringlePRL35} revealed an enhanced emission
of neutrons along the fission direction of unknown origin.
While the evidence did not permit any definite conclusion,
it was speculated that those might be scission neutrons.
A few years later, Franklyn \etal\ \cite{FranklynPLB78}, 
also analyzing measured $n$-$n$ correlations,
found that a 20\% component of scission neutrons was present,
but they saw no indication of enhanced polar emission.

Samant \etal\ \cite{SamantPRC51} compared measured differential distributions
of prompt neutroms from $^{235}$U($n_{\rm th}$,f) 
with the calculated contributions of evaporation from the fragments
and found that the neutron distribution has a CM-centered component
at the level of $\approx$10\%, increasing somewhat with TKE.
Based on analyses of three independent experiments
carried out by different methods, Kornilov \etal\ \cite{Kornilov2000} 
also concluded that around 10\%\ of the neutrons from $^{252}$Cf(sf)
are due to the scission process.
More recently, Kornilov \etal\ \cite{Kornilov2007} measured the prompt
neutron spectrum from $^{235}$U($n$,f) induced by neutrons of energy 0.5\,\MeV\
and found that 25\% of the yield arises from a central source
of somewhat higher effective temperature.
Moreover, a neutron surplus of $\approx$30\% was found 
in the transverse direction; 
these neutrons represent $\approx$10\%  of the prompt fission neutron yield
and could not originate from fully accelerated fragments.
Concurrently, Gagarski \etal\ \cite{Gagarski2008} measured neutron-neutron
angular correlations in $^{252}$Cf(sf) and found that 
about 10\%\ of the prompt neutrons were emitted isotropically
and therefore tentatively interpreted as scission neutrons.

However, a later comprehensive review \cite{Capote2015} concluded that
a very small percentage of neutrons emitted in fission are scission neutrons
(whose energy can be several tens of MeV),
while an overwhelming fraction of the prompt neutrons
are evaporated from the accelerated fission fragments.
Furthermore, Serot \etal\ \cite{SerotND2016},
having augmented the {\sc fifrelin} fission event generator \cite{FIFRELIN}
with a schematic scission neutron source,
found that a 2\% scission neutron component yielded
a very good agreement with the Mannhart evaluation \cite{Manhart1987}.
Recently, Vorobyev \etal\ \cite{Vorobyev2020a,Vorobyev2020b,VorobyevEPJW239},
comparing calculated evaporation spectra with experimental data,
found an excess at 90$^\circ$ and estimated the yield of scission neutrons
to be 2.5\% -- 4.5\% for $^{233}$U($n_{\rm th}$,f), $^{235}$U($n_{\rm th}$,f), 
$^{239}$Pu($n_{\rm th}$,f), and $^{252}$Cf(sf).

Finally, very recently, Schulc \etal\ \cite{Schulc2023,SchulcPRC109},
using multiple neutron dosimetry reactions with very high thresholds,
observed that very energetic neutrons are emitted in fission.  Earlier,
the $^{90}$Z($n,2n$) neutron dosimetry evaluation with threshold at 12\,\MeV\
 was used to adjust the high energy component of the prompt fission neutron 
spectrum from $^{235}$U(n$_{th}$,f) \cite{Trkov2015,Trkov2015a} 
which was evaluated at the IAEA \cite{Capote2015} and was subsequently 
adopted in the ENDF/B-VIII.0 evaluation \cite{Capote2018,Brown2018}. 
The measurements by Schulc \etal\ \cite{Schulc2023,SchulcPRC109}
are in excellent agreement with that evaluation.


Alongside the experimental efforts,
scission neutron production has been studied theoretically.
As already suggested by Anderson \etal\ \cite{AndersonPR55},
Bohr and Wheeler \cite{BohrWheelerPR56} also concluded that 
prompt neutrons may be produced already at the time of the division process,
as well as being evaporated from the fragments.
Later on, as a general consideration, Halpern \cite{Halpern1965} suggested that
the sudden release of potential energy associated with the neck rupture
could cause emission of scission neutrons, but no specific model was developed.

The rapid change in the nuclear mean field 
accompanying the rather sudden change
from a connected to a disconnected configuration is expected 
to agitate the neutrons in the spatial region between the fledging fragments.
Modeling the potential as a one-dimensional square well 
with a growing bump in the center, Fuller \cite{FullerPR126} 
estimated that some fraction of the prompt fission neutrons
are produced by this mechanism and, depending on the assumptions,
they will emerge with kinetic energies of several MeV.

This mechanism was subsequently studied to various degrees of refinement
by Carjan, Wada, and collaborators
\cite{CarjanNPA792,Rizea:2008xzn,CarjanPRC82,RizeaCPC9,CarjanPRC85,RizeaNPA909,
Wada:2013pmc,Rizea:2013vnq,Wada:2015xja,Carjan:2015zja,Rizea2025,CarjanPLB747,
CapotePRC93,Wada:2017hxq,Rizea:2018mqt,Wada:2018pty,CarjanPRC99,
CarjanIJMPE28,Wada:2013jcu}
who calculated the effect on the initial neutron wave functions
both in the sudden approximation and with more realistic shape evolutions,
including absorption and scattering by the receding fragments.
The characteristics of the resulting scission neutrons
were found to be rather sensitive to the assumptions made
about the specified evolving potential.
In particular, the scission neutron multiplicity was found to range 
from one tenth to one third of the total prompt neutron multiplicity,
and their differential distribution was found to vary from being
``similar'' to those evaporated from the accelerated fragments
to being ``completely different'' from those.
Such a significant sensitivity suggests that considerable information about 
the scission dynamics could be gained by more detailed neutron measurements.


The mechanism considered in the present investigation comes into play 
just {\em after} the fissioning system has completed its division
into two separate but still distorted fragments.
For each of the fragments, 
the relaxation of the shape towards its equilibrium form
generally agitates the individual nucleons,
thereby converting the potential energy associated with the distorted shape
into additional intrinsic excitation.
In particular, a sufficiently fast inwards surface motion
may cause some nucleons to become unbound
and thus possibly be emitted.

This mechanism for emission of neutrons in association with scission was
first suggested about 40 years ago by M{\"a}dler \cite{MaedlerZPA321}.
He argued that reflection from an inwards-moving surface of speed $U$ 
should boost the normal speed of a nucleon from $v_\perp$ to $v_\perp+2U$,
which may render it unbound.
For example, if $U$ is about 10\%\ of the Fermi speed
then a nucleon in the Fermi surface would, on average,
gain about 10\,\MeV\ upon reflection.

Naming this the 'catapult' mechanism,
M{\"a}dler used the Time-Dependent Hartree-Fock treatment
to study it for one-dimensional (\ie\ slab) configurations.
These calculations suggested that the high-energy tail of the emitted catapult 
neutrons would stand out clearly against the background from the neutrons
evaporated by the equilibrated fragments.
Furthermore, he conjectured that the effect would be largest for fission events
with low fragment kinetic energy because the associated scission configurations
are more elongated and, consequently, 
the fledging fragments are initially more distorted.

Recently, Abdurrahman \etal\ \cite{Abdurrahman:2024mxk,Abdurrahman:2023uzd}
studied the emission of scission neutrons with the
Time-Dependent Density Functional Theory (TDDFT) \cite{TDDFT}.  
Their analysis of the results for $^{252}$Cf(sf) \cite{Abdurrahman:2024mxk}
suggested that there are two distinct types of scission emission:
An immediate transverse emission from the neck region and
a later parallel emission from the far ends of the two receding fragments.
We suspect that the early type arises from a mechanism
of the type studied in Refs.\ \cite{FullerPR126,
CarjanNPA792,Rizea:2008xzn,CarjanPRC82,RizeaCPC9,CarjanPRC85,RizeaNPA909,
Wada:2013pmc,Rizea:2013vnq,Wada:2015xja,Carjan:2015zja,Rizea2025,CarjanPLB747,
CapotePRC93,Wada:2017hxq,Rizea:2018mqt,Wada:2018pty,CarjanPRC99,
CarjanIJMPE28,Wada:2013jcu}.
As for the later type,
the authors of Ref.\ \cite{Abdurrahman:2024mxk} identify it with
the catapult mechanism suggested by M{\"a}dler \cite{MaedlerZPA321}
(and they introduce the alternate term ``slingshot'' for it).
A subsequent study \cite{Abdurrahman:2023uzd}
included also fission of $^{236}$U and $^{240}$Pu
which yielded very similar results.
For the three cases together, the scission neutrons had energies up to
$16-18$\,\MeV, with a mean energy of $\approx$3\,\MeV, and
their multiplicity was conservatively estimated to be $9-14$\% of the total.

The TDDFT treatment presents the state-of-the-art framework for 
self-consistent fully microscopic nuclear dynamics.
However, it is rather computer demanding which makes it impractical 
to carry out detailed follow-up studies \cite{Ibrahim}.
For the purpose of providing guidance for further experimental 
and theoretical investigations,
the present study explores the catapult mechanism more thoroughly
by describing the nucleonic motion in terms of classical trajectories
inside the distorted fission fragments.

\section{Physical scenario}

The evolution of the fissioning nuclear system leads to scission 
where the neck connecting the two fledging fragments snaps. 
Numerical simulations based on self-consistent and parameter-free models,
such as those in Ref.\ \cite{Abdurrahman:2023uzd},
suggest that the configurations shortly afterwards resemble 
two coaxial pear-like shapes with the thin ends facing each other.
Being localized around the common symmetry axis, those protrusions
are the remnants of the neck and they quickly shrink as a result of the 
associated cost in surface energy, leading to more gently deformed fragments.

\begin{figure}[bht]	    
\includegraphics[ trim={4cm 6cm 3cm 7cm}, clip, width=0.45\textwidth]
		  {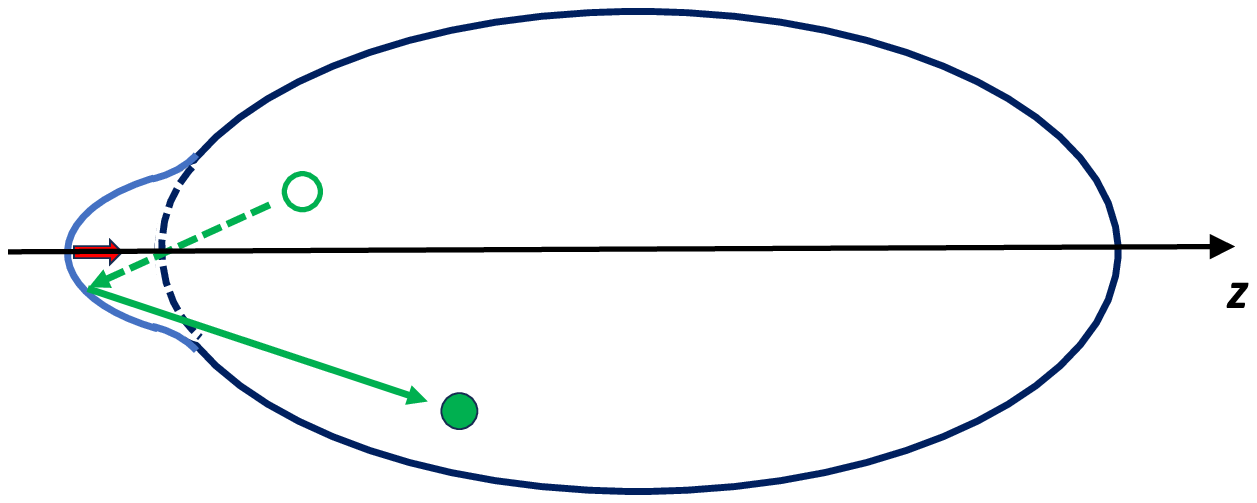} 
\caption{\label{f:shape}
Shortly after scission the remnant of the ruptured neck
endows each fragment with a bulge that is rapidly shrinking
towards the smooth reference shape.
An approaching nucleon is reflected from the inwards moving bulge surface,
thereby being boosted to a higher energy.}
\end{figure}

When an individual nucleon inside the fragment arrives at 
the shrinking protrusion, it is being reflected by the inwards moving surface
and boosted to a higher energy, possibly becoming unbound,
as first suggested by M{\"a}dler \cite{MaedlerZPA321}.
The overall scenario is illlustrated in Fig.\ \ref{f:shape}.

\subsection{Bulge geometry}

In the domain of the bulge, \ie\ near the rear end of the fragment,
the smooth reference shape, towards which the bulge shrinks,
is assumed to be approximately spheroidal,
in accordance with the microcopic calculations \cite{Abdurrahman:2023uzd}.
In cylindrical coordinates, the reference profile $(\rho_s,z_s)$ 
is given by $\rho_s^2/a^2+z_s^2/c^2=1$, where the axis ratio $c/a$ is specified.
(The standard value is $c/a=1.8$,
corresponding to the shape shown in Fig.\ \ref{f:shape}.)

\begin{figure}[tbh]	    
\includegraphics[trim={8.5cm 8cm 9.5cm 6.8cm},clip,width=0.4\textwidth]
		{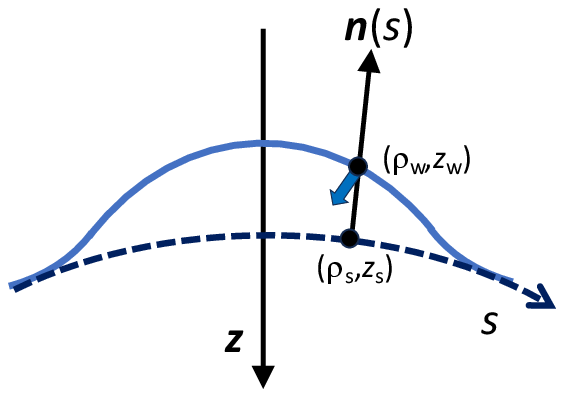}    
\caption{\label{f:bulge}
Illustration of the bulge geometry:
The profile of the smooth reference surface is shown by the dashed black curve
along which the distance $s$ is measured.
The bulge profile is shown by the solid blue curve;
it has a Gaussian shape when the displacement from $(\rho_s,z_s)$,
is measured along the local normal $\bld{n}(s)$.
The wide blue arrow shows the normal velocity of the local surface element 
of the shrinking bulge at $(\rho_w,z_w)$.}
\end{figure}

The bulge surface is defined by the function $\h(s)$,
where $s$ is the arc length along the reference profile,
giving the outwards displacement from the reference surface
along the local normal $\bld{n}(s)$, as illustrated in Fig.\ \ref{f:bulge}.
Judging from the microscopic simulations \cite{Abdurrahman:2023uzd},
the profile of the bulge appears to be somewhat bell shaped
and it will here be approximated by a Gaussian at any time,
\beq
\h(s;t) = \h_0(t) g(s),\,\,\ g(s)=e^{-s^2/2\sigma_0^2}\ .
\eeq
The decreasing bulge height is given by $\h_0(t)$, 
while the parameter $\sigma_0$ controls the width of the bulge 
which is assumed to remain constant as the bulge shrinks.
The standard values, $\h_0(0)$\,=\,2\,fm and $s_0$\,=\,1\,fm,
yield initial bulge shapes that roughly match those from
the microscopic calculations \cite{Abdurrahman:2023uzd}.

Relative to the reference shape,
the presence of the bulge increases the surface area of the fragment
by an amount $\Delta S(\h_0)$ and the associated potential energy is
$V_{\rm bulge}(\h_0)=\gamma\Delta S(\h_0)$, 
where the nuclear surface tension is $\gamma\approx0.9\,\MeV/\fm^2$ 
for a typical fission fragment.

\subsection{Bulge dynamics}

For the fragment, 
it is as if its surface had been pulled out locally and then released
(hence the term `slingshot' introduced in Ref.\ \cite{Abdurrahman:2023uzd}).
Thus, when the neck snaps, the involved part of the surface 
is suddenly being accelerated inwards by the driving force
$F_{\rm drive}(\h_0)=-\del V_{\rm bulge}/\del\h_0$.

The induced inwards surface motion 
is resisted by a friction force, $F_{\rm frict}$, 
which may be estimated from the familiar one-body wall formula \cite{wall}
whose predictions agree well with the results obtained microscopically
with TDDFT \cite{BulgacPRC100}.
The dissipation rate associated with the deforming shape is thus taken as
\beq
\dot{Q}_{\rm bulge} = m\rho\bar{v}\! \int\! U(s)^2 d^2\sigma 
= K(\h_0) \dot{\h}_0^2 .
\eeq
Here $m$ is the nucleon mass, 
$\rho$ is the nucleon density in the bulk region of the fragment,
and $\bar{v}\approx\mbox{$3\over4$}v_F$ is the mean nucleon speed
(with $v_F\approx8.4\,\fm/10^{-22}{\rm s}$ being the Fermi speed).
The integral is over the deforming part of the surface,
with $U(s)=\dot{\h}_0 g(s) \cos\beta(s)$ 
being the velocity of the bulge surface element
in the direction normal to the bulge
($\beta$ is the angle between the normal to the bulge surface
and the normal to the reference surface).
The friction coefficient $K(\h_0)$ can then be calculated
and the friction force is given by $F_{\rm frict} = -K(\h_0)\dot{\h}_0$.

Because the one-body dissipation is rather strong \cite{wall}
the surface motion is strongly damped and its time evolution 
can be obtained by balancing the driving force against the friction force,
$F_{\rm drive}+F_{\rm frict}\doteq0$.
So the rate of change of the bulge height is 
$\dot{\h}_0 = F_{\rm frict}(\h_0)/K(\h_0)$.
With the standard values, $\h_0(0)=2\,\fm$ and $\sigma_0=1\,\fm$,
the initial central velocity is $\dot{\h}_0(t=0)\approx-0.2\,v_F$.

\subsection{Nucleon trajectories}

Those nucleons that happen to be reflected from the inwards-moving bulge surface
will be boosted to a higher energy.
This study estimates the number and energy of neutrons
emitted by this ``catapult' mechanism in fission.

{\em Reflections from the shrinking bulge.}
The rate of nucleons arriving at a surface element is given by \cite{wall}
\beq
{\cal N} = \int_{v_\perp>U(s)}(v_\perp-U(s)) f(\bld{p}) {d^3\bld{p}\over h^3}
\approx \half \bar{v}_\perp \rho .
\eeq
Here $f({\bld p})=1/[1+\exp((E-E_F)/T)]$ 
is the phase-space occupancy in the nuclear bulk region,
with the kinetic energy being $E=\half m v^2$. Furthermore, 
$\bar{v}_\perp$ is the average of the nucleon velocity in the direction 
normal to the surface element, $v_\perp={\bld v}\cdot\hat{\bld n}$.
For a Fermi-Dirac distribution at moderate temperatures, $T\ll E_F$,
the nucleon flux amounts to
${\cal N}\approx \half \bar{v_\perp}\rho
\approx0.27/(10^{-22}{\rm s}\,{\rm fm}^2)$.
About 61\% of those are neutrons for a typical fission fragment.

\begin{figure}[tbh]	        
\includegraphics[ trim={4cm 6cm 2cm 7cm},clip,width=0.45\textwidth]
		  {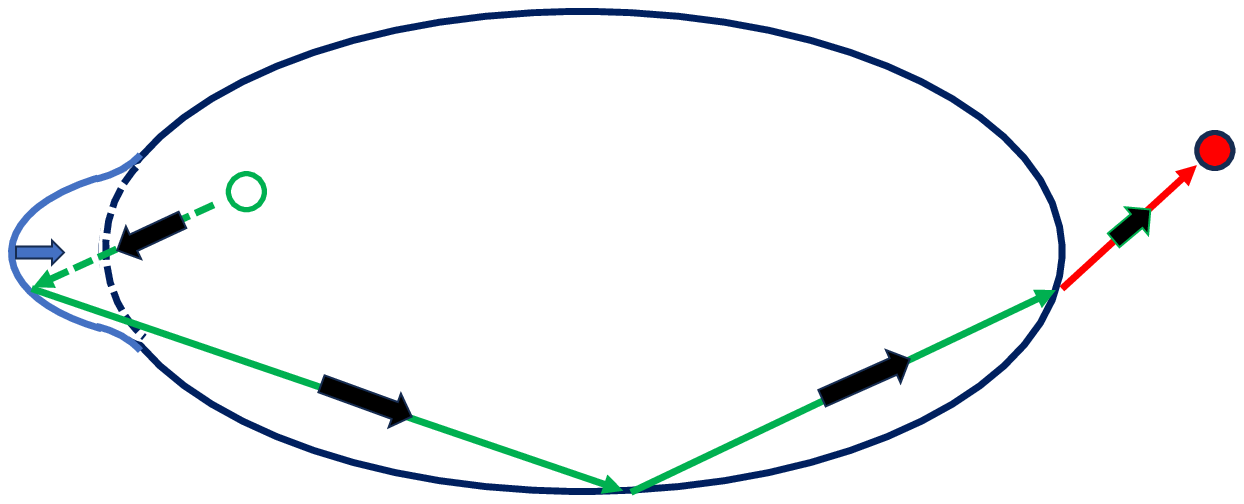} 
\caption{\label{f:path}
Illustration of the trajectory of a catapult neutron:
Arriving from the bulk region, 
it is reflected from the inwards moving bulge surface, becoming unbound,
and then starts moving  through the fragment.
When encountering the surface it is either reflected
(if its local normal motion is insufficient) or emitted
along a refracted trajectory.}
\end{figure}		     	

The reflection of an outward moving neutron from the inward moving surface
increases its speed because its normal velocity relative to tbe moving 
surface is being reversed.
The associated energy increase is given by
$\Delta E =-2mUv_\perp+2mU^2$.
(Both terms are positive because the surface velocity is negative, $U<0$.)
Usually the surface speed is smaller than the Fermi speed $v_F$
and the first term dominates.

In order to eventually escape from the nucleus,
the boosted neutron must have a kinetic energy 
that exceeds the escape threshold energy, $E_{\rm esc}=E_F+S_n\approx44\,\MeV$,
where $S_n$ is the neutron separation energy in the primary fragment.
(Hence, the multiplicity and energy of catapult neutrons
should be largest for those fragments that have the lowest separation energy.)

{\em Propagation and emission.}
After its reflection from the shrinking bulge,
the boosted neutron reenters the bulk region of the fragment
and propagates through its mean field.
Because nucleons are fermions, their direct interactions with other nucleons
are strongly suppressed by Pauli blocking of the possible final states, 
thus endowing them with a long mean free path 
and allowing then to occupy well-defined single-particle orbitals.

\begin{figure}[tbh]	        
\includegraphics[trim={2cm 8cm 3cm 6.5cm},clip,width=0.55\textwidth]
			   {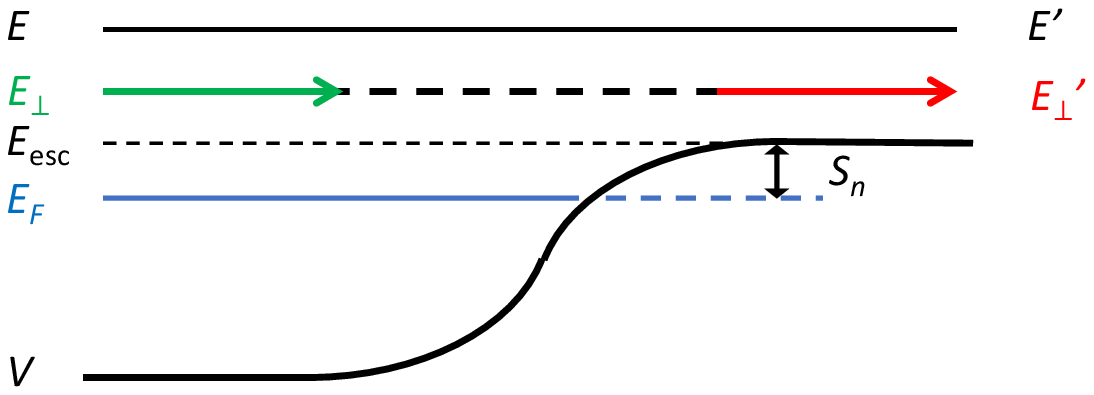}	
\caption{\label{f:emit}
Illustration of the emission process:
At the nuclear surface the effective potential $V$ increases 
from its bulk value to zero.
An unbound neutron has a total energy $E$ 
above the escape threshold $E_{\rm esc}=E_F+S_n$ and
approaches the surface with the normal energy $E_\perp=\half mv_\perp^2$.
If $E_\perp$ also exceeds $E_{\rm esc}$ the neutron is emitted
with a reduced normal energy, $E_\perp'=E_\perp-E_{\rm esc}$,
while its motion parallel to the surface remains unchanged.}
\end{figure}		     	

Because of its long mean free path, a boosted neutron is likely
to encounter the nuclear surface (several times)
before it experiences a collision with another nucleon
and the calculated results depend only weakly 
on the value employed for the long mean free path.

Most often, when encountering the nuclear surface,
the kinetic energy of the neutron in the direction normal
to the local surface element, $E_\perp=\half mv_\perp^2$,
 is smaller than the escape energy $E_{\rm esc}$ and it will 
merely reflect off the surface and continue 
moving through the nucleus towards its next wall encounter.
However, occasionally it has suficient normal energy to escape the
nuclear potential well, $E_\perp>E_{\rm esc}$, and it will be emitted,
with an appropriately reduced normal energy $E_\perp'=E_\perp-E_{\rm esc}$.
The associated refraction of its trajectory
tends to divert the emitted neutrons away from the fragment axis.

The propagation and emission of a catapult neutron is illustrated in
Figs.\ \ref{f:path} and \ref{f:emit}.

\section{Calculated results}

For the physical scenario described above,
the emission of catapult neutrons has been studied
by following the fate of the neutrons 
that reflect off the inwards moving bulge wall.

Over the course of the bulge shrinkage,
these simulations typically sample 50 million test neutrons 
from the Fermi-Dirac distribution in the considered fragment,
assumed to have a temperature of $T=1\,\MeV$,
with the reflecting surface element
being sampled randomly from the bulge surface for each neutron.

After the initial reflection, a few per cent of the boosted neutrons
have an energy above the escape threshold, $E>E_{\rm esc}$,
and their subsequent trajectories are followed as they move through the fragment.
Whenever such a neutron arrives at the fragment surface,
its energy in the direction of the local normal is calculated
and it is emitted if that exceeds $E_{\rm esc}$.
Otherwise it is reflected elastically and its further trajectory is followed.

\begin{table}[bth]		 
\begin{tabular}{c|ccc}\hline\hline\\[-2ex]
$ \overline{\nu}_{\rm cat}\,:\,\overline{E}_{\rm cat}$
& $\sigma_0\!=$0.8\,\fm & $\sigma_0\!=$1.0\,\fm & $\sigma_0\!=$1.2\,\fm \\[1ex]
\hline\\[-2.5ex]
$\h_0\!$=1.6\,\fm~ &~   2.8: 11.5 ~&~ 3.2: 8.9  ~&~ 3.6: 7.3 \\
$\h_0\!=$2.0\,\fm~ &~   3.2:  12.0 ~&~ 3.8: 9.1  ~&~ 4.2: 7.4 \\ 
$\h_0\!=$2.4\,\fm~ &~ \,3.7: 12.4 ~&~ 4.3: 9.3  ~&~ 4.8: 7.5 \\
\hline\hline
\end{tabular}
\caption{\label{table} The expected number of catapult neutrons per fragment
$ \overline{\nu}_{\rm cat}$ (times 100) 
together with their mean kinetic energy $\overline{E}_{\rm cat}$ (in MeV)
calculated for various initial bulge heights $\h_0$ 
and bulge widths $\sigma_0$, for fragments having $c/a=1.8$.}
\end{table}

Table \ref{table} summarizes how the main results
depend on the initial bulge profile.
The average multiplicity of catapult neutrons and their mean energy
are shown for variations of the initial bulge height and width
around the adopted standard values ($\h_0=2\,\fm$ and $\sigma_0=1\,\fm$).
It is apparent from the table that a narrower bulge 
leads to a smaller catapult yield but
a larger catapult energy (because the bulge profile grows steeper).
Furthermore, an increase of the initial bulge height
increases both the yield and the mean energy, as would be expected.

Both the yield of catapult neutrons and their energy
are moderately dependent on the reference shape of the fragment.
Thus, in the typical well-deformed region,
an increase of $c/a$ from 1.8 to 2.0 increases
$\overline{\nu}_{\rm cat}$ 3.2\% to 3.3\% and
$\overline{E}_{\rm cat}$ from 9.14 to 9.66 \MeV,
while a decrease of $c/a$ to 1.6 leads to 
$\overline{\nu}_{\rm cat}=3.0\%$
and $\overline{E}_{\rm cat}=8.65\,\MeV$.
A spherical shape yields
$\overline{\nu}_{\rm cat}=2.7\%$
and $\overline{E}_{\rm cat}=7.7\,\MeV$.

The results in Table \ref{table} are for a single fragment.
Thus, for example,
for fission of $^{236}$U into a deformed and a spherical fragment,
one would expect $0.038+0.027=0.065$ catapult neutrons,
corresponding to $\approx$2.7\%\ of the total prompt neutron multiplicity
$\overline{\nu} = 2.42$ \cite{Carlson2018}.

\SKIP{
Thus, if both fragments have $c/a=1.8$, the expected number 
of catapult neutrons per fission event is about $2\times 0.038 = 0.076$.
For $^{235}$U($n_{\rm th}$,f), which has a mean prompt neutron multiplicity 
of $\overline{\nu}\approx2.42$ \cite{Carlson2018},
that would correspond to about $3.1\%$ of the total,
for the standard bulge profile.}

\begin{figure}[tbh]	    
\includegraphics[trim={2cm .8cm 7cm 8cm},clip,width=0.33\textwidth,angle=-90]
			   {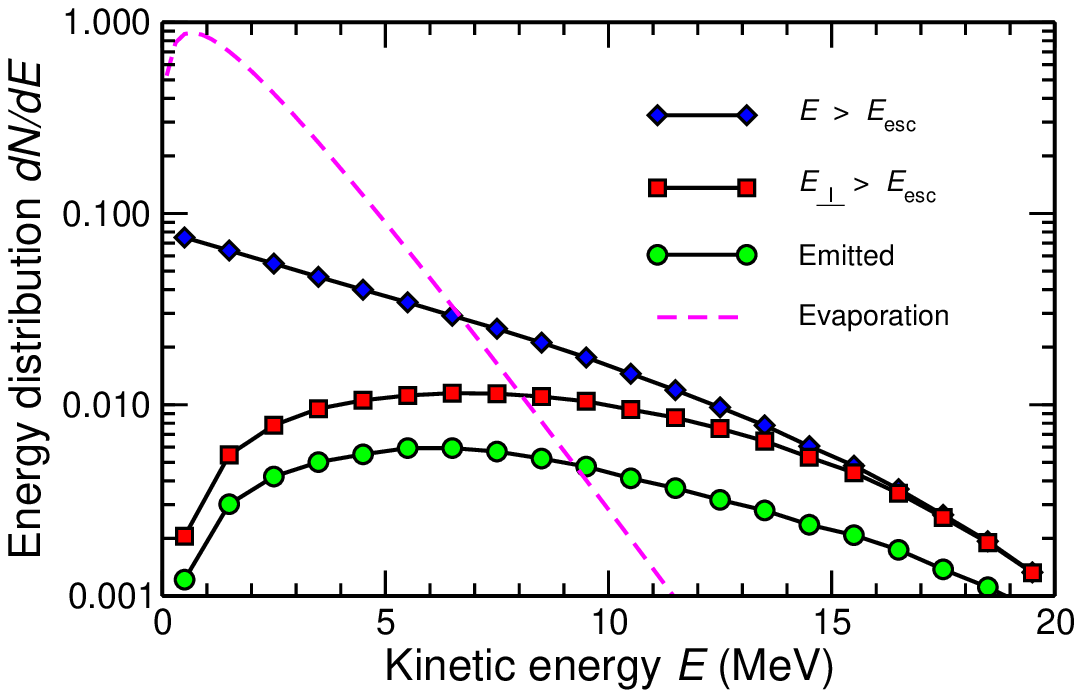}	
\caption{\label{f:dNdE}
The energy spectrum of three classes of neutrons 
after their reflection from the inwards moving bulge surface:
{\em solid blue diamonds:} those that are unbound,
\ie\ their energy exceeds the escape threshold, $E>E_{\rm esc}$;
{\em solid red squares:} 
those that leave the bulge surface with a normal energy above the threshold,
 $E_\perp>E_{\rm esc}$;
{\em solid green circles:} those that have a local normal energy above 
$E_{\rm esc}$ when subsequently encountering the fragment surface
and therefore actually escape;
the contributions from both fission fragments have been combined.
The catapult emissions from both fragments have been combined
and the evaporation spectrum for $^{235}$U($n_{\rm th}$,f) 
calculated with {\sc freya} is also shown (dashed curve).}
\end{figure}

Figure \ref{f:dNdE} shows the energy distribution
of the boosted neutrons at the various stages of the process.
After their initial reflection off the inwards moving bulge surface,
those neutrons that have become unbound 
(\ie\ their energy exceeds the escape energy, $E>E_{\rm esc}$) 
have a hard high-energy tail which around $10\,\MeV$ can be characterized
by temperature of $\approx6\,\MeV$.
Though unbound, most of those neutrons (about 85\%) will not be emitted
because they have insufficient normal energy 
when subsequently encountering the nuclear surface.
To give an impression of this important feature,
Fig.\ \ref{f:dNdE} shows the energy distribution
of those neutrons whose normal energy right after their initial reflection
exceeds the escape threshold, $E_\perp > E_{\rm esc}$.
About one third of those neutrons succeed in escaping
during one of their subsequent encounters with the fragment surface
and the spectral distribution of these catapult neutrons is also displayed.

The present calculations were using a nuclear temperature of $T=1\,\MeV$,
which is a typical temperature at scission.
However, the partitioning of the excitation energy between the fragments
varies significantly \cite{PLB803}
and the fragment temperatures typically deviate from the scission temparature.
A 20\%\ increase in $T$ causes the calculated catapult multiplicity 
to increase by about 3\%, so the $T$ dependence is quite weak.

Figure \ref{f:dNdE} also shows the energy spectrum of the neutrons
evaporated from the fragments in the reaction $^{235}$U($n_{\rm th}$,f)
for which the average neutron multipliciy is $\overline{\nu}\approx2.4$.
as obtained with the fission event generator {\sc freya} \cite{FREYA}.
Although the catapult neutrons appear only at the level of $\approx3\%$,
their spectrum is significantly harder 
and they dominate over the evaporation neutrons above $\approx9.3\,\MeV$.

\section{Concluding remarks}

The catapult mechanism is universal, 
operating in both induced and spontaneous fission
and it is not expected to exhibit much energy dependence.
Indeed, it is present not only in any fledging fission fragment 
but also, more generally, in any nucleus having been prepared 
-- by any means -- with a local surface bulge.

The present exploratory calculations suggest that the dependence
on the specific fragment is relatively moderate, 
though some general trends appear.
Thus, for a given initial bulge height, 
a wider bulge increases the multiplicity but hardly affects the energy,
while a higher initial bulge, for a given bulge width,
increases both the multiplicity and the energy.
This finding supports the speculation by M{\"a}dler \cite{MaedlerZPA321}
that fission events with lower TKE (and hence a more elongated neck)
will produce more energetic catapult neutrons.

In conclusion, the present study suggests that catapult neutrons exist,
at the level of a few per cent,
and they have energies far in excess of typical evaporation neutrons,
a key characteristic that should make it easier to identify them experimentally.
The calculations support the conclusion of
Schulc \etal\ \cite{Schulc2023,SchulcPRC109} 
that very energetic neutrons are emitted during fission, and their finding 
that the measured neutron spectrum dominates over the standard evaporation 
spectrum above $\approx$10\, \MeV\ 
is consistent with the results shown in Fig.\ \ref{f:dNdE}.

\section*{Acknowledgments}
The authors wish to thank Ibrahim Abdurrahman for helpful discussions 
and for making the time evolution of the TDDFT density contours available.
Partial support by the IAEA for JR and RV is gratefully acknowledged.
This work was also supported in part by the U.S.\ Department of Energy
under Contracts No.\ DE-AC02-05CH11231 (JR) and DE-AC52-07NA27344 (RV).

\SKIP{
~\\[1.5ex] {\bf \centerline{DATA AVAILABILITY}}

~\\[-1.5ex]
The data that support the findings of this article 
are openly available \cite{data}.
}



                        \end{document}